\documentclass[aps,prx,twocolumn,groupedaddress,showpacs,floatfix,superscriptaddress,longbibliography]{revtex4-2}
\usepackage[plainpages=false,pdfpagelabels,colorlinks=true,linkcolor=red,urlcolor=blue,citecolor=blue,pdftitle={Title},pdfauthor={},pdfdisplaydoctitle=true,pdfduplex=DuplexFlipLongEdge]{hyperref}
\usepackage{epsfig}
\usepackage{amsmath,amssymb}
\usepackage{physics}
\usepackage{bm}
\usepackage{graphicx}
\usepackage[dvipsnames,usenames]{color}
\usepackage[normalem]{ulem}
\usepackage{soul}
\usepackage{notes2bib}
\usepackage{xcolor}

\bibnotesetup{
note-name = ,
use-sort-key = false
}

\begin{document}

\title{The SU(N) Holstein Model}
\author{Chunhan Feng} 
\affiliation{Max Planck Institute for the Physics of Complex Systems, Nöthnitzer Straße 38, 01187 Dresden, Germany}
\author{Linh Pham}
\affiliation{Department of Physics and Astronomy, University of California, Davis, CA 95616, USA}
\author{George Batrouni} 
\affiliation{Universit\'e C\^ote d'Azur, CNRS, Institut de Physique de Nice (INPHYNI), 06000 Nice, France}
\author{Richard T. Scalettar}
\affiliation{Department of Physics and Astronomy, University of California, Davis, CA 95616, USA}

\date{\today}


\begin{abstract}
From the condensed matter physics perspective, the most natural single 
orbital tight-binding Hamiltonians, and hence the most widely studied, contain two fermionic species,
corresponding to spin up and spin down electrons.  
In cold atom systems, however, SU(N) symmetry, in which $N > 2$ fermionic
species reside within a single band, also occurs.  In order to understand such
experiments, the SU(N)
Hubbard model has been increasingly studied.  Here we present determinant Quantum Monte 
Carlo simulations of the SU(N) {\it Holstein} Hamiltonian, in which $N$ fermionic
species couple to a single local phonon mode.  We show that at half filling 
it has an insulating charge density wave phase (CDW) at low temperatures, in which empty sites
alternate with sites with $N$ particles.  We determine the $N=3$ CDW phase diagram in the 
temperature, $T$, versus electron-phonon coupling, $\alpha$, plane at fixed phonon frequency $\omega_0$
and half-filling $\rho=1.5$.
The critical temperature $T_c$ for $N=3$ can be as high as twice the maximum attainable
for $N=2$.
We also obtain the $N$ dependence of $T_c$ for a representative, fixed, $\omega_0$ and $\alpha$.
\end{abstract}

\maketitle

\section{Introduction}\label{sec:Intro}

The study of SU($N$) quantum magnetism historically originated from the
mathematical technique of large-$N$
expansions\cite{read83,affleck85,bickers87,affleck1988,read1989some,read1989,wu06,auerbach12}.
Even as $N$ increases, quantum fluctuations remain
important since SU($N$) symmetry prevents spins from becoming
classical\cite{auerbach12,wu06}.
Much attention has been devoted to the Heisenberg limit of {\it localized} spins
where exotic phenomena including three-sublattice magnetic ordering are found to occur
\cite{toth10,bauer12,nataf14,corboz11,hermele11,romen20,yamamoto20}.

Theoretical models with SU($N$) symmetry have
been discussed in connection with
the exact SU($N$) nuclear spin symmetry\cite{wu06,cazalilla09,gorshkov10,cazalilla14}
which is realized in
fermionic isotopes of alkaline-earth-metal-like atoms (AEAs).
Here the study of {\it itinerant} magnetism, as described by the SU($N$) Hubbard
Hamiltonian is paramount.
Recent experiments on ultracold atomic gases have emphasized the feasibility of
realizing the model
\cite{wu2003,honerkamp2004ultracold,gorshkov2010,taie2012SU,pagano2014one,cazalilla14,hofrichter2016,taie2022observation,tusi2022flavour,pasqualetti2024}, and the possibility of another platform- artificial lattices of dopant-based quantum dots in silicon- is also being explored\cite{das2011hubbard,salfi2016quantum,wang2022experimental,liu2022zoo,wang2024efficient}.
As for $N=2$, the SU($N$) Heisenberg model emerges as the strong
coupling limit of the SU($N$) Hubbard model for $N>2$, connecting
the itinerant and localized limits.

Study of the SU($N$) Hubbard model with quantum simulations has provided insight
into exotic magnetic ordering patterns and the formation of long
range order in the absence of nesting
\cite{taie2022observation,padilla2023metal,feng2023metal}.
However, such studies are
challenging because of a sign problem\cite{loh90,troyer05,mondaini2022quantum}
which dramatically worsens as $N$ increases.

Here we explore the generalization of the Holstein model\cite{holstein59}
to $N>2$ fermionic species.  One motivation is to attain a better understanding
of two dimensional SU($N$) systems in which a finite temperature transition
can occur:  in the Holstein model, at half filling, a discrete Ising
symmetry is being broken in the charge density wave phase,
unlike the Heisenberg and Hubbard cases where in $d=2$ the breaking of
a continuous spin symmetry is possible only at $T=0$\cite{mermin1966absence}.

To address these, and related, questions,
we begin in Sec.~\ref{sec:MandM} with a description of the
Holstein model with general $N$ and the (straightforward) extension of the
Determinant Quantum Monte Carlo (DQMC) algorithm for its
simulation.
Section~\ref{sec:Results} provides data for key observables
in the SU($3$) Holstein model, leading to the construction of its phase diagram
in the plane of temperature and electron-phonon coupling
at fixed (half-filled) density and phonon frequency.
We also explore the $N$ dependence of $T_c$ for a fixed parameter set in order
to determine the asymptotic behavior at large number of species.
Machine Learning (ML) provides a rapidly developing set of methodologies for
phase detection in classical\cite{wang2016discovering,hu2017discovering} and quantum models\cite{carrasquilla2017machine,costa2017principal,ch2017machine,dong2019machine,johnston2022perspective}.
Section \ref{sec:ML} provides complementary analysis of the SU(3)
Holstein model using ML approaches.
Finally, Sec.~\ref{sec:Conclusions} presents a summary as well  as 
discussions of SU($N$) symmetry in the electron-phonon systems
and the question of superconductivity in the doped lattice.

\section{Model and Methods}\label{sec:MandM} 

\vskip0.07in \noindent
The Holstein Hamiltonian\cite{holstein59} is
one of the most simple tight-binding descriptions of the
electron-phonon (el-ph) interaction.
\begin{align} \label{eq:Holst_hamil}
\hat {\cal H}_{\rm Holstein} =
-& t 
\sum_{\langle ij \rangle, \sigma}
\big(\hat c^{\dagger}_{i \sigma}
\hat c^{\phantom{\dagger}}_{j \sigma} +
\hat c^{\dagger}_{j \sigma}
\hat c^{\phantom{\dagger}}_{i \sigma}
 \big)
-  \mu \sum_{i \sigma}  \hat n_{i \sigma}
\nonumber \\
+& \alpha \sum_{i \sigma}
\hat n_{i\sigma}
\hat{X}_{i}
+ \, \frac{1}{2M} \sum_{ i }
\hat{P}^{2}_{i}
+ \frac{M}{2} 
\omega_{ 0 }^2
\sum_{ i } \hat{X}^{2}_{i}
\,\,,
\end{align}
Here a collection of fermionic degrees of freedom with creation (destruction) operators $c^{\dagger}_{i \sigma}\,\,
(c^{\phantom{\dagger}}_{i \sigma})$,
labeled by spatial site $i$ and flavor index $\sigma = 1, 2, \cdots N$,
hop on near-neighbor sites $\langle {i,j} \rangle$ of a square lattice of
linear size $L$.  These interact with
oscillator degrees of freedom $\hat X_{i},
\,\hat P_{i}$ which satisfy the usual canonical commutation relations and which are localized on each lattice site with electron-phonon
coupling $\alpha$.
Particle-hole symmetry implies $\rho=N/2$ (i.e.~half filling) at $\mu = -N \alpha^{2}/(2 \, \omega_{0}^{2})$
for any temperature on a bipartite lattice.
We report results in
terms of the dimensionless coupling $\lambda_D \equiv \alpha^2/(W\,\omega_{0}^2)$,
where the non-interacting bandwidth $W = 8\,t$,
and will focus mostly on the case $N=3$.
We choose our units by setting $M=1$ and $t=1$.

We will explore the properties of  Eq.~\ref{eq:Holst_hamil} with determinant
Quantum Monte Carlo (DQMC)\cite{blankenbecler1981monte,white89,cohen2022fast}.
In this method, the partition function is written as a path integral
by discretizing the imaginary time $\beta = L \Delta \tau$ (we typically take $\Delta\tau=0.1$).
In this construction
the phonon operators are replaced by a phonon field $x(i,\tau)$ in $2+1$ dimensions,
and give rise to a `bosonic' contribution to the action,
$S_{\rm bose} \equiv \frac{1}{2} \Delta \tau \sum_{i,\tau} 
\{ \, \omega_{0}^{2} \, x(i,\tau)^{2}
  + [\,(x(i,\tau+1) - x(i,\tau))/\Delta \tau\,]^{2} \,\}$ \cite{creutz1981statistical}.

Because the fermions appear quadratically, they may be traced out analytically yielding a product of $N$ determinants in the Boltzmann weight.  All
fermionic species couple in the same way to the phonon field, so these determinants
are identical.  The DQMC algorithm for $N>2$ then proceeds in the standard way used for $N=2$:
changes are proposed to the phonon field at individual space-imaginary time
points. The changes to $S_{\rm bose}$ and the fermion determinants are then
used for a Metropolis accept/reject decision.
These local moves are supplemented by global changes to $x(i,\tau)$
for a fixed spatial site $i$ and all imaginary time
slices $\tau$, tuned in such a way as to cross over the energy
barrier associated with empty and doubly occupied configurations\cite{cohen2022fast}.

\begin{figure}[t]
    \centering
    \includegraphics[width=3.7in,height=3.0in]{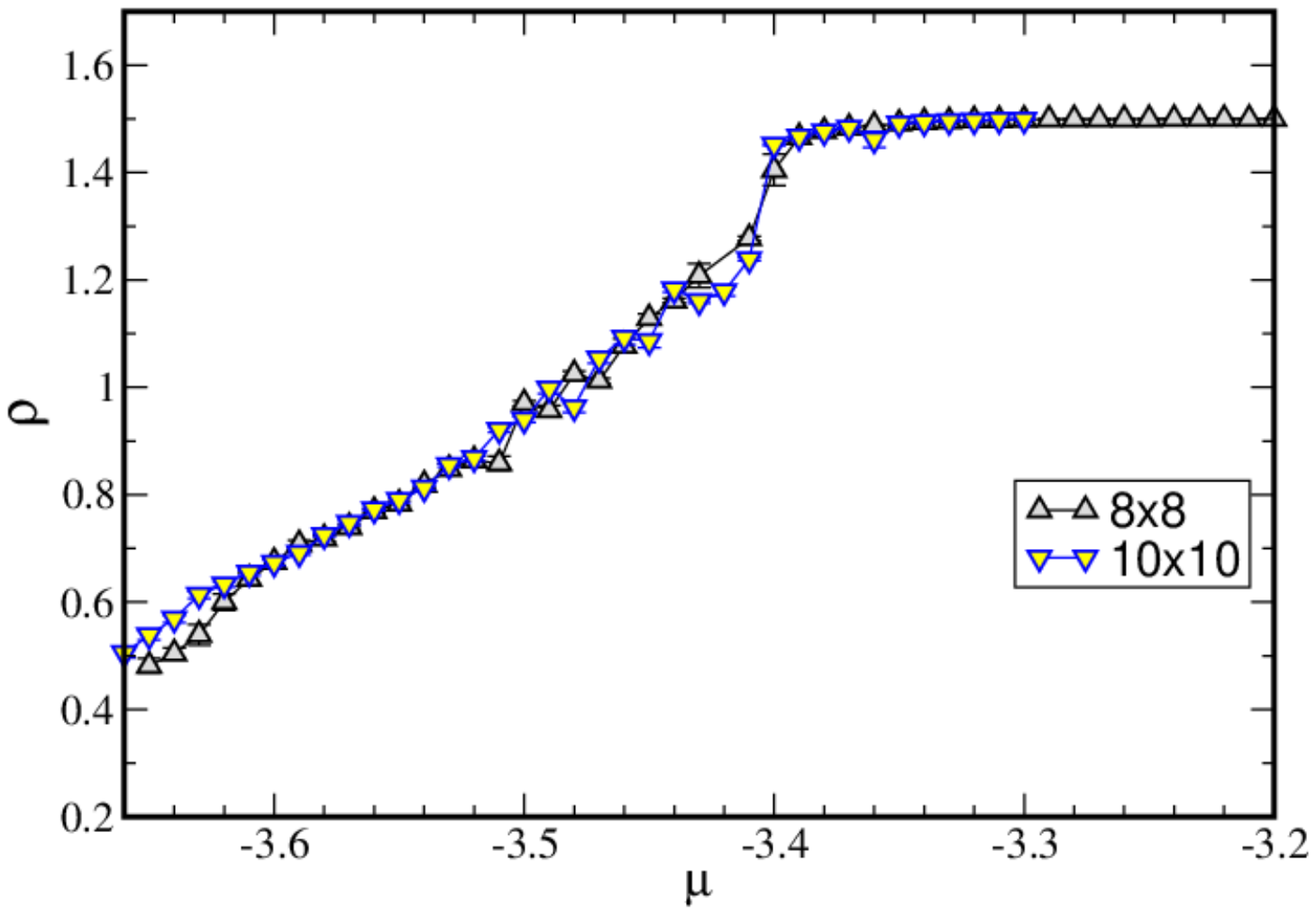} 
    \caption{Density $\rho$ as a function of chemical potential $\mu$ at inverse temperature $\beta \, t=12$, $\omega_0=1$, $\alpha=\sqrt{2}$, $\lambda_D=0.25$. Particle-hole symmetry implies the chemical potential for half filling is $\mu = -N \alpha^{2}/(2\omega_{0}^{2}) = -3$ for the current parameter values.
    The plateau in $\rho$ indicates the presence of an incompressible insulating CDW phase. The gap is symmetric with respect to $\mu=-3$ and is therefore $\Delta \sim 0.8t$. 
    We note a small discontinuous jump in $\rho$ as the system approaches half filling. This indicates a first order transition and is similar to that observed in the case of two fermion flavors\cite{bradley2021superconductivity}.
    }
    \label{fig:rhovsmu}
\end{figure}

QMC studies of
$\hat {\cal H}_{\rm Holstein}$
have uncovered many interesting features,
from polaron and CDW formation to (`conventional')
superconductivity\cite{scalettar89,marsiglio90,vekic1992,freericks93,li2015a,li2015b,weber2018,costa2018phonon,esterlis18,hohenadler19,chen2019,zhang2019,cohenstead20,feng2020, feng2020_2, dee20,bradley2021superconductivity,nosarzewski21,zhang2022, araujo22,ly2023comparative, kvande2023} (SC).
SC is, however, challenging to observe,
since it requires very low temperatures\cite{bradley2021superconductivity}.
One reason is the on-site fermionic pairs which form due to the phonon-mediated attractive interaction
need to break apart and reform in order to move.  The polaron mass is large, suppressing SC order
\cite{Sous2018light,Li20}.

Since the determinants are identical, it follows that for even $N$, DQMC for the SU($N$) Holstein model is manifestly free of the sign
problem.  Interestingly, we have found no sign problem for $N$ odd at the parameter ranges studied here.  The 
phonon kinetic energy operators $\hat P_{i}^2$ 
in the Hamiltonian lead to a contribution in the
bosonic part of the action which suppresses large fluctuations of
the phonon field in imaginary time.  These inhibit the 
occurrence of negative determinants, since the determinant
of an imaginary time independent phonon field must be positive. 
We emphasize that this absence of the sign problem for odd $N$ is restricted to parameters where the system is away from the anti-adiabatic limit, $\omega_0 \rightarrow \infty$, because in that limit the model maps onto the attractive Hubbard model with an odd number of flavors, which does suffer from the sign problem.

\section{Results}\label{sec:Results}

We begin by determining the dependence of the total density $\rho$ 
on chemical potential $\mu$ in Fig.~\ref{fig:rhovsmu}.  There is a clear plateau at half-filling, $\rho=1.5$.  This vanishing of the compressibility is a signal of an insulating CDW phase which we confirm below to originate in the alternation of triply occupied and empty sites. 
We also note a small discontinuous jump in $\rho$ as the system approaches half-filling. 
This indicates a first order transition and is similar to that observed in the case of $N=2$ fermion flavors\cite{bradley2021superconductivity}. Data for two lattice sizes, $8\times8$ and $10\times10$ are in good agreement, as is expected for local observables like the density. 
We also observe patterns in the spin-spin correlations and in $\rho(\mu)$ which suggest the possibility of magnetic order at integer filling $\rho=1$.
However, the data are noisy, likely due to competition between different local minima, and it
appears important to choose lattice sizes for which the order is not frustrated, e.g.~$L=6,12$
which allow both for spatial period two or three\cite{feng2023metal,padilla2023metal}. We leave a detailed investigation of this regime for future studies.

We can see that this half-filled insulating phase has long range charge
order by measuring the associated structure factor,
\begin{align}
S_{\rm cdw} = \frac{1}{L^{2}} \sum_{ij} (-1)^{i+j} \langle \, \hat n_i \hat n_j \, \rangle
\label{eq:scdw}
\end{align}
where $\hat n_{i} = \sum_{\sigma} \hat n_{i\sigma}$ is the total
density of all species $\sigma$ on site $i$ and the phase factor
$(-1)^{i+j}$ alternates in sign between the two sublattices
of the bipartite square lattice.

\begin{figure}[t]
    \centering
    \includegraphics[width=3.7in,height=2.5in]{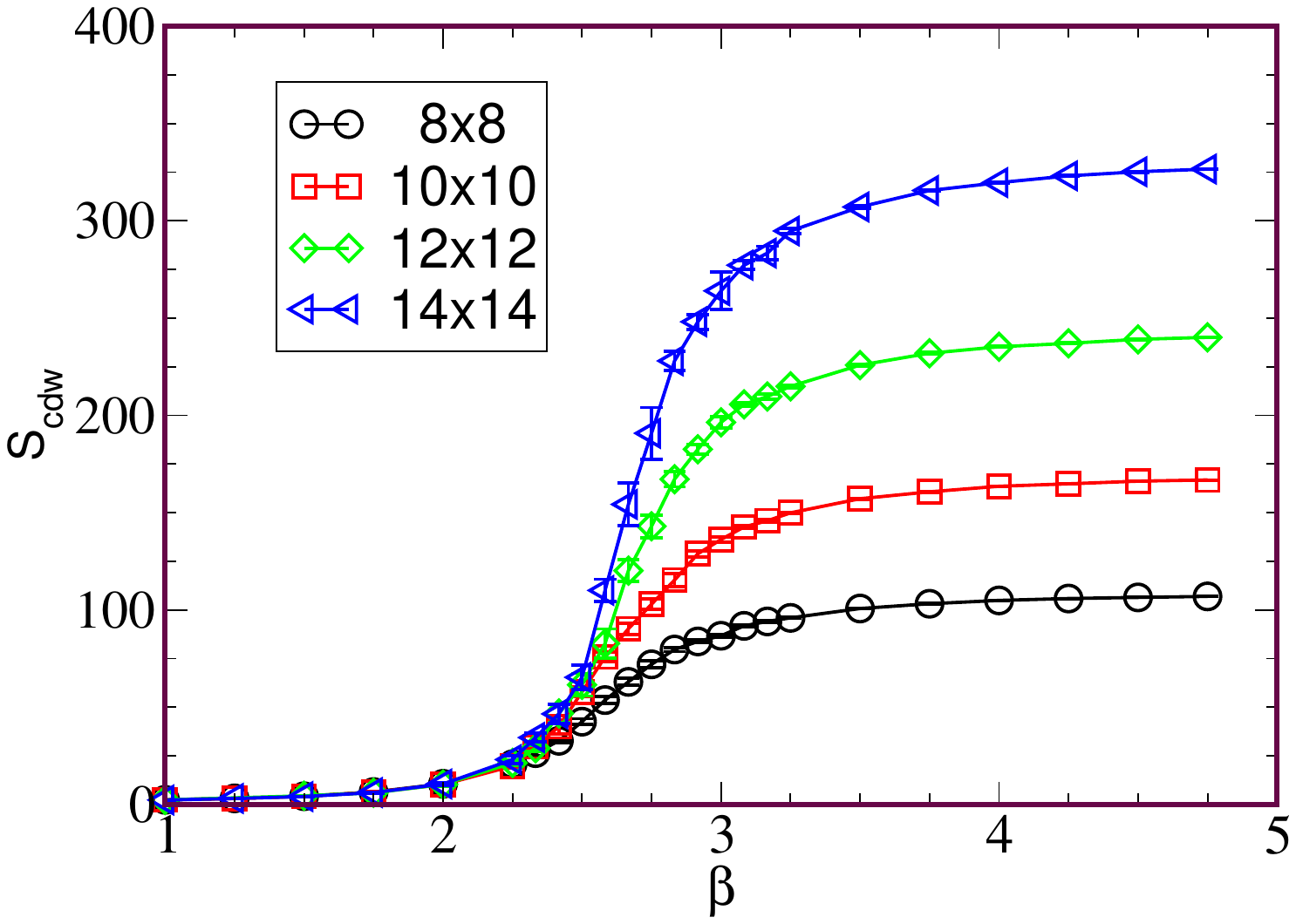} 
    \caption{
    The CDW structure factor $S_{\rm cdw}$ is shown as a function
    of inverse temperature $\beta \, t$ at half-filling and 
     $\lambda_D=0.3025, \, \omega_0=t$ 
    for four linear lattice sizes $L$ and $N=3$ fermionic species.  
    A critical $\beta_c$ can be roughly inferred from
    the value at which $S_{\rm cdw}$ grows with $L$.
    }
    \label{fig:scdw}
\end{figure}

In a disordered, metallic, phase, the real space charge correlations
$ \langle \, \hat n_i \hat n_j \, \rangle $ will fall off exponentially
with separation $| i-j| $ and $S_{\rm cdw}$ will be independent
of lattice size $L$.  However, in a phase with long range order
the double sum over all pairs of sites $i,j$ will lead to
$S_{\rm cdw} \propto L^{2}$.
For high $T$ (small $\beta$), Fig.~\ref{fig:scdw} shows the former behavior,
while for low $T$ (large $\beta$), the charge structure factor grows with $L$.
For a perfectly ordered phase we have $S^{\rm max}_{\rm cdw} = L^{2} \frac{N^2}{4}$ 
because $\frac{1}{4}$ of the pairs of sites lie on the occupied sublattice.
For $N=3, L=10$ this yields $S^{\rm max}_{\rm cdw} = 225$.
The data of Fig.~\ref{fig:scdw} do not saturate this bound, indicating
the presence of substantial quantum fluctuations due to the 
fermionic hopping $t$, even in the ground state at $T=0$.

\begin{figure}[t]
    \centering
    \includegraphics[width=3.7in,height=2.5in]{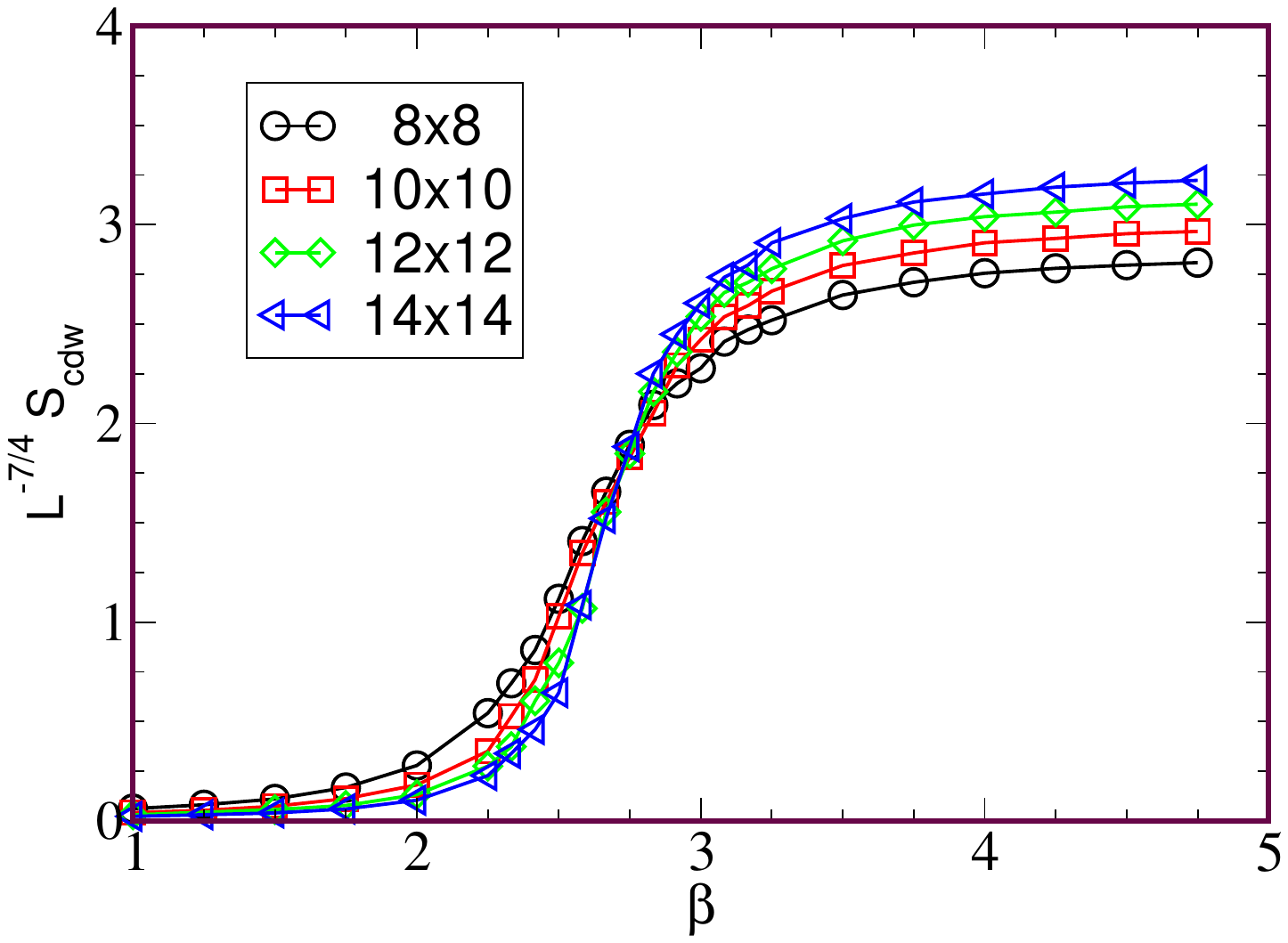} 
    \caption{
    The scaled CDW structure factor $L^{-7/4} S_{\rm cdw}$ is shown as a function
    of inverse temperature $\beta \, t$ at half-filling for the same parameters
    as Fig.~\ref{fig:scdw} ($\lambda_{D}=0.3025, \omega_0=t$).
    The crossing at $\beta_c \, t \sim 2.75$ ($T_c \sim 0.36\, t$) gives
    the location of the CDW transition.
    }
    \label{fig:fss}
\end{figure}

The range of inverse temperature $2 \lesssim \beta \lesssim 3$ 
at which $S_{\rm cdw}$ begins to acquire 
a size dependence in Fig.~\ref{fig:scdw} gives a rough estimate of the
critical point.  Finite size scaling, as shown in Fig.~\ref{fig:fss}, provides
a precise determination.  Here we scale the structure factor with
$L^{-\gamma/\nu}$ using the $d=2$ Ising values appropriate to breaking
a ${\cal Z}_{2}$ symmetry.  A crossing is observed at inverse temperature
$\beta_{c} \sim 2.75$ (critical temperature $T_{c} \sim 0.36\,t$),
consistent with the range 
$2 \lesssim \beta_{c} \lesssim 3$  noted above.

\begin{figure}[t]
    \centering
    \includegraphics[width=3.7in,height=3.0in]{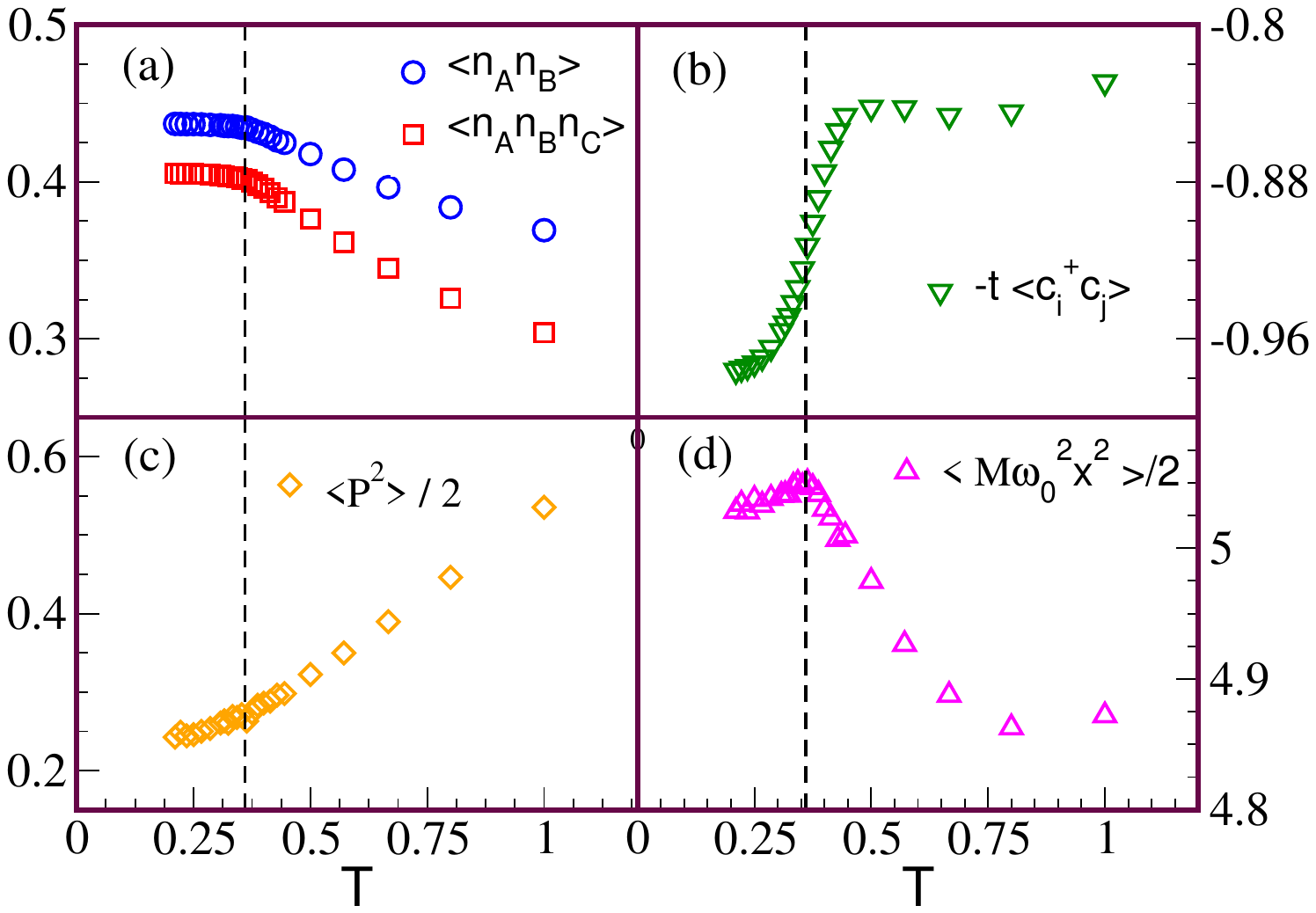} 
    \caption{
   Behavior of local observables as a function of temperature T for an $L=10$ lattice
   and $\lambda_{D}=0.3025$.  
   (a)  The expectation values of two and three fermion occupations.  The latter corresponds to the
   fraction of triply occupied sites, while the former gets contributions from both double and triple occupations;
   (b)  electron kinetic energy;
   (c)  phonon kinetic energy;
   (d)  phonon potential energy.
   The vertical dashed line in each panel is the CDW transition temperature determined
   by the crossing of the scaled structure factor (Fig.~\ref{fig:fss}).
    }
    \label{fig:other}
\end{figure}

The transition to the CDW insulator is reflected also in other (local) observables.
    Figure \ref{fig:other} gives the density of triply occupied sites, the electron kinetic energy,
    and the phonon kinetic and potential energy as the temperature is varied.
    For decreasing $T$, the phonon-mediated attraction causes
    an increase in sites with three fermions, so that in the ground state almost half the 
    sites are triply occupied.  The electron kinetic energy shows a sharp increase in magnitude
    at $T_{\rm cdw}$, reflecting the greater ability of fermions to hop when occupied sites
    are surrounded by empty ones.  While the phonon kinetic energy exhibits no clear signal
    of the transition, the phonon potential energy has a sharp rise as $T$ is decreased-  
    the phonon displacement grows as the fermions become bound.
    The origin of the subsequent fall-off is obscure, but a similar phenomenon is observed in the
    temperature evolution of the local moment in the 2D Hubbard model, 
    which, at strong coupling $U$,  similarly first rises 
    as the fermions cool, but then shows a smaller decrease as long range antiferromagnetic 
    order is established 
    across the lattice\cite{paiva2001signatures,Schafer2021,Feng2025}.

\begin{figure}[t]
    \centering
    \includegraphics[width=3.5in,height=2.7in]{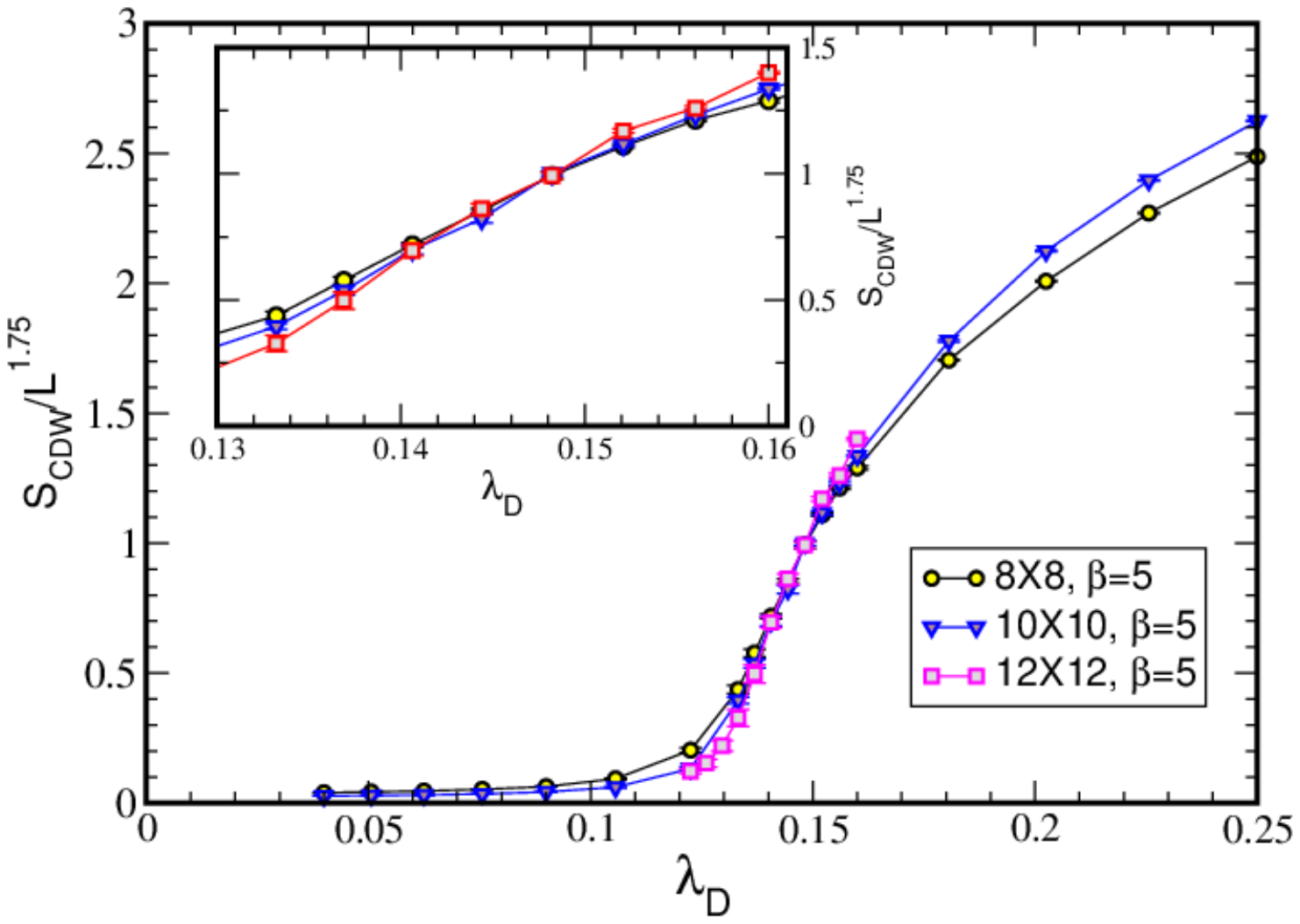} 
    \caption{
    Scaled structure factor $L^{-\gamma/\nu}S_{\rm cdw}$
    as a function of $\lambda_{D}$ for fixed inverse temperature $\beta=5$ and linear lattice
    sizes $L=8, 10, 12$.    
    Inset: close up of the crossing region.
    Critical exponents for the 2D Ising universality class $\gamma/\nu=7/4$ were used. $\omega_0=1$.
    }
    \label{fig:lambdaDsweep}
\end{figure}

In addition to entering the ordered 
CDW phase by decreasing the temperature at fixed $\lambda_{D}$, as in 
Fig.~\ref{fig:scdw}, one can also increase $\lambda_{D}$ at fixed 
(low) $T$.  Such a sweep is shown in 
    Fig.~\ref{fig:lambdaDsweep} for $T=0.2\,t$ for the
    scaled structure factor.   A crossing is observed at $\lambda_{D} \sim 0.14$.

\begin{figure}[t]
    \centering
    \includegraphics[width=3.5in,height=2.7in]{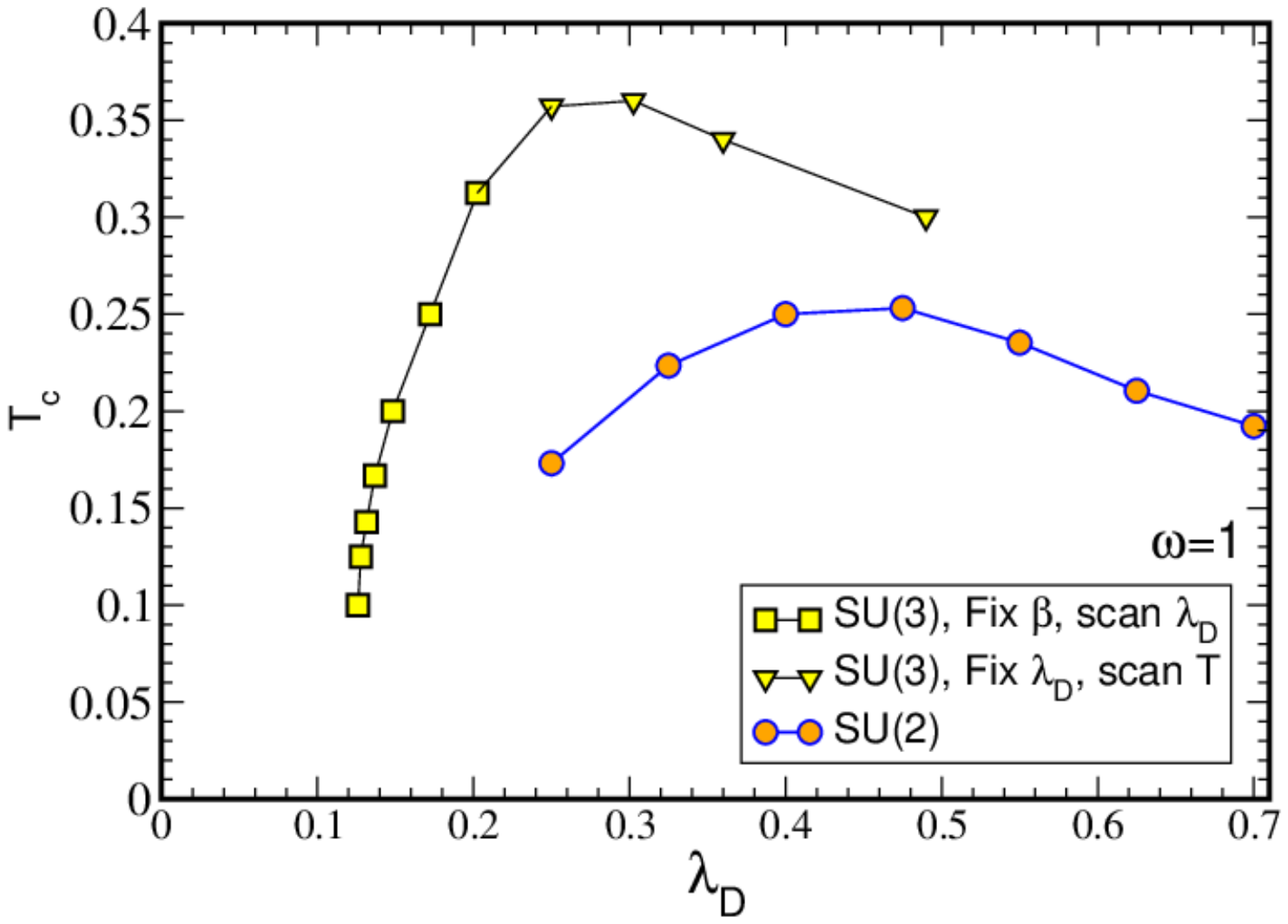} 
    \caption{
    The phase diagram of the SU($N$) Holstein model in the 
    temperature-dimensionless coupling plane at fixed frequency
    $\omega_{0}=1$ and $N=3$.
    Results from `vertical' sweeps (changing temperature)
    such as Fig.~\ref{fig:fss} are denoted
    by downward triangles.
    Results from `horizontal' sweeps (changing $\lambda_{D}$)
    such as Fig.~\ref{fig:lambdaDsweep} are denoted by
    squares.
    Values for the conventional $N=2$ case are given for comparison
    (circular markers)\cite{issa2025learning}.
    }
    \label{fig:phasediagram}
\end{figure}

We have performed similar calculations 
    for a range of dimensionless coupling
    constants $\lambda_{D}$ at constant phonon frequency 
    $\omega_{0}=1$.
The resulting critical temperatures of the $N=3$ Holstein model are shown in 
    Fig.~\ref{fig:phasediagram} 
    as a function of $\lambda_{D}$.
    Results from sweeps of $\lambda_{D}$ at fixed $T$ are used at weak coupling, where the
    phase boundary rises nearly vertically.
    Results from sweeps of $T$ at fixed $\lambda_{D}$ are used at larger coupling, where the
    phase boundary is more horizontal.
    The low values of $T_{c}$ for small $\omega_{0}$ are well-known:
    at weak coupling one expects $T_{c} \sim \omega_{0} \, e^{-1/\lambda_{D}}$ \cite{zhang19}.
    The more gradual fall-off at large $\lambda_{D}$ reflects a breakdown of
    Migdal-Eliashberg theory \cite{esterlis18}.

\begin{figure}[t]
    \centering
    \includegraphics[width=3.5in,height=2.7in]{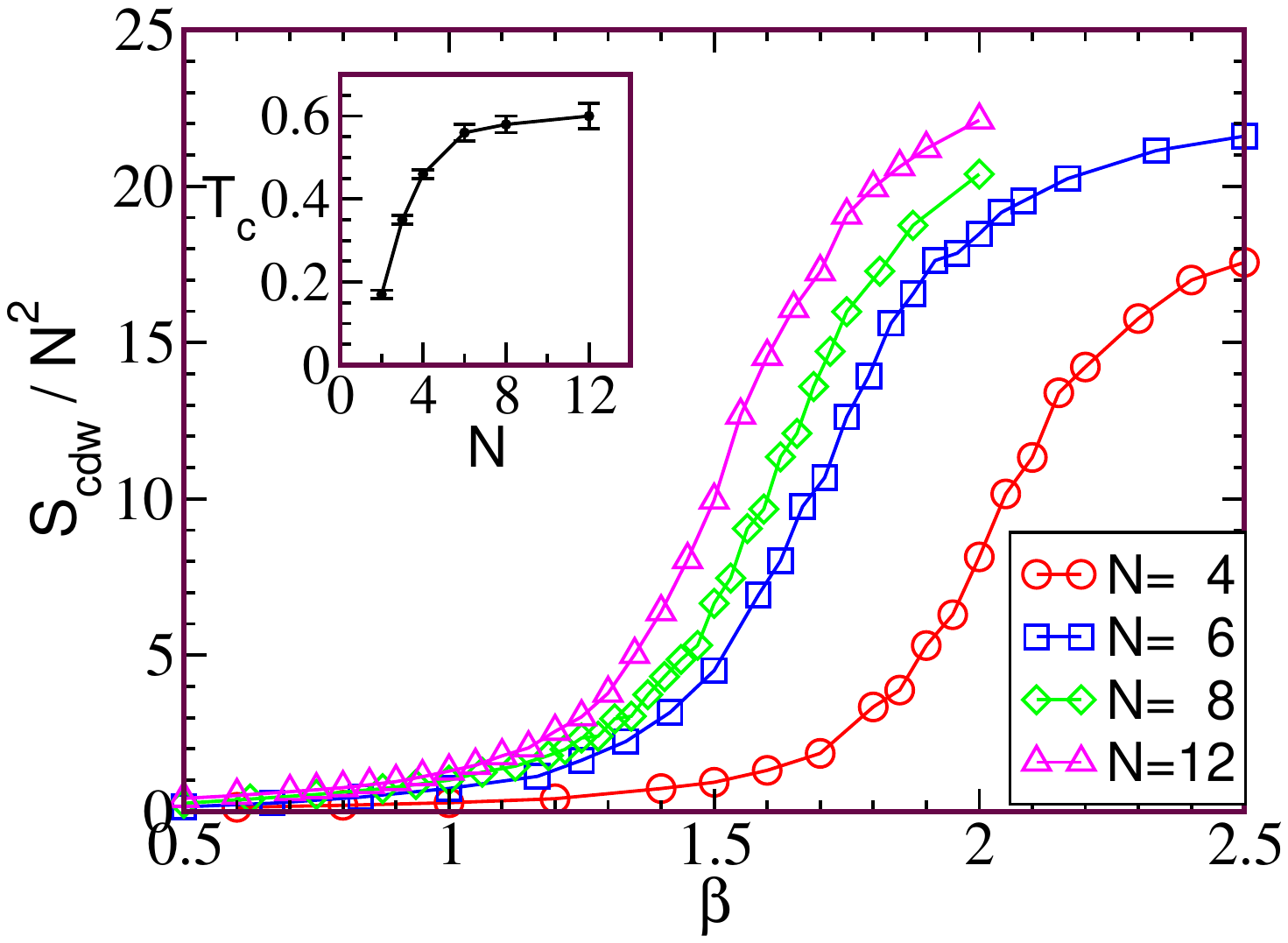} 
    \caption{
    The charge structure factor on a $10\times10$ lattice is
    shown as a function of inverse temperature $\beta$.  We have used $\lambda_{D}=0.25$ and
    $\omega_{0}=1$ and
    have normalized $S_{\rm cdw}$ by $N^{2}$.  (See text.)
    The inset gives the critical temperature $T_c(N)$ obtained from crossing plots for
    different lattice sizes.  $T_c$ values are given for $N=2,3,4,6,8,12$ but
    $S_{\rm cdw}$ is shown only for the latter four values, since $N=3$ data
    are given earlier in this paper, and $N=2$ data in the literature.
    }
    \label{fig:TcofN}
\end{figure}

Finally, we show in Fig.~\ref{fig:TcofN} a plot of the CDW structure factor 
$S_{\rm cdw}$ on a $10\times10$ spatial lattice
as a function of $\beta$ for different $N$.  As noted earlier, the magnitude of
$S_{\rm cdw}$ is expected to
have a `trivial' $N^2$ dependence associated with the presence of $N$ fermions on
the sites of the occupied sublattice.  We have normalized $S_{\rm cdw}$
by $N^2$ for this reason.  As we already saw in comparing the critical temperatures for $N=2$ and $N=3$
(Fig.~\ref{fig:phasediagram}), the inverse temperature at which
charge correlations develop decreases as $N$ increases, emphasizing that
CDW order occurs at higher $T_{c}$.
The inset to Fig.~\ref{fig:TcofN} shows $T_c \rightarrow 0.63\,t$ asymptotically
at large $N$.

Another feature of Fig.~\ref{fig:TcofN} is that $S_{\rm cdw}$ saturates 
at larger values in the ground state (large $\beta$) as $N$ grows, even
after the normalization by $N^2$.
This also has an analog in the Hubbard model where the antiferromagnetic
structure factor at $T=0$ increases with $U$.  Large interactions reduce the
quantum fluctuations which remain even when thermal fluctuations
are turned off in the ground state ($T=0$).
Perfect CDW order would have $S_{\rm cdw}/N^{2} = L^{2}/4 = 25$ (for $L=10$).
Figure \ref{fig:TcofN} shows the approach to this limit as both $\beta$ and $N$ 
become large.

\begin{figure}[t]
    \centering
    \includegraphics[width=3.5in,height=2.7in]{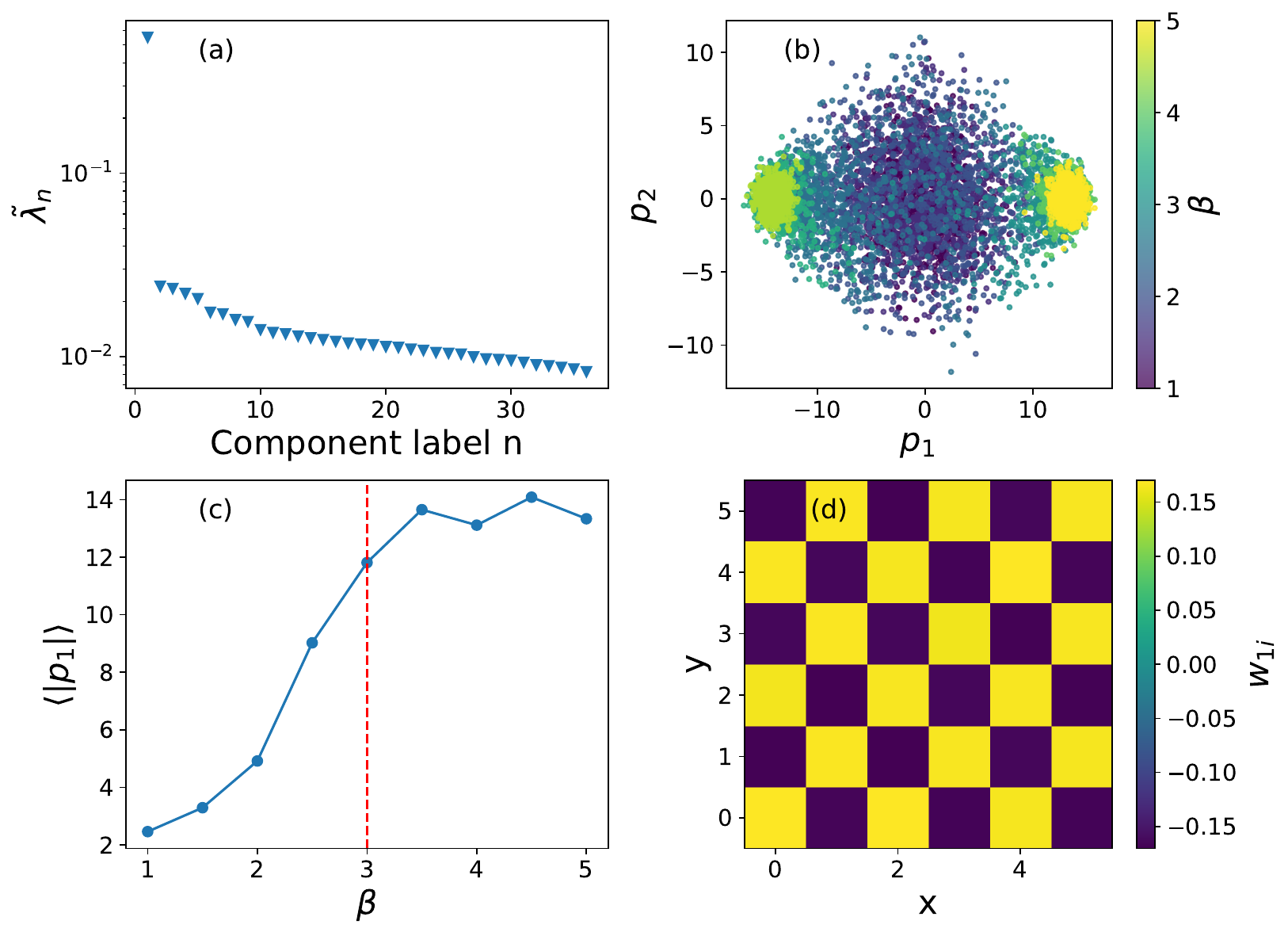} 
    \caption{
    (a)  Relative variances $\tilde \lambda_n$ of the PCA eigenvalues;
    (b)  Scatter plot of the projections $p_1$ and $p_2$ of the snapshots onto the first two
    PCA eigenvectors;
    (c)  The quantified first component $p_1$ as a function of inverse temperature $\beta$.
    The vertical dashed line gives the location, $\beta_c \sim 3$, of the transition obtained from finite size
    scaling of the structure factor;
    (d)  The first weight vector. Calculations are performed on a  $6 \times 6$ lattice with $\omega=1$ and $\alpha = 1.70$. 
    }

    \label{fig:PCA}
\end{figure}

\section{Machine Learning Analysis}\label{sec:ML}

One can also examine the CDW transition in the SU(N) Holstein model using Machine
Learning methods.  Here we apply the most straight-forward option, principal
component analysis (PCA), which has previously been used for the $N=2$ Holstein model\cite{costa2017principal}.
In this method, multiple configurational snapshots are retained from our quantum Monte Carlo
simulation at each of a collection of inverse  temperature values bracketing the phase transition.
There are different choices for the snapshots, e.g.~measurements of the local fermionic
Green's function $G_{\sigma}(r,r)$ whose value gives the density $\rho_{\sigma}(r)$, or
the phonon configuration $x(r,\tau_{0})$ at a particular imaginary
time slice $\tau_0$\footnote{One could also take snapshots at all imaginary time slices.
However this does not significantly affect the results.}.  
Any choice of $\tau_0$ is equivalent owing to the periodic structure in
imaginary time.  We have chosen the latter option, phonon snapshots, which provide
a somewhat more crisp picture of the physics.

The snapshots are then placed in an array $X(r,j)$, where 
$r=1,2,\cdots N$ labels the spatial site.
$j$ labels the snapshot and takes on values $1,2, \cdots N_{s} N_{\beta}$ where $N_s$ is the number of samples at each of
$N_{\beta}$ inverse temperatures.  The ($N$ dimensional square) matrix $X X^{T}$ is then diagonalized.
PCA works well when the resulting eigenspectrum is dominated by a few large values.
The overlaps $p_i$ (dot products) of each snapshot with the associated eigenvectors are computed.  
Using the first two overlaps, a scatter plot of $(p_1,p_2)$ then shows a characteristic
change in topology from a single blob to two distinct concentration regions when 
$\beta > \beta_c$ in situations when a ${\cal Z}_2$ symmetry is broken, as in the
present case.

Figure \ref{fig:PCA} shows the results.  
Panel (a) gives the eigenspectrum structure. 
Specifically, the relative variances 
 $\tilde{\lambda}_{n} \equiv \, \lambda_n / \sum_{n}\lambda_{n}$ 
 are computed from the eigenvalues $\lambda_{n}$ of $X X^{T}$.
 These measure the proportion of the total data variance which can be attributed to
 each PCA component.
They fall rapidly with index $n$, emphasizing that the data lie in a low dimensional subspace.
and that the behavior of the system is dominated by a small number of collective variables.
The scatter plot of panel (b) indicates the expected bifurcation of the distribution
at a $\beta_c \,t \sim 3$ consistent with that inferred from Fig.~\ref{fig:fss}.
The first principal component $p_1$ behaves almost like an order parameter (panel (c)), growing
rapidly at $\beta \gtrsim \beta_c$. Finally, the eigenvector of maximal eigenvalue reflects
the spatial structure of the CDW order, as seen in panel (d).

\section{Conclusions}\label{sec:Conclusions}

Over the last decade,
considerable numerical work has been done on the SU($N$) Hubbard 
Hamiltonian\cite{zhou2016,xu2018,unukovych2021,singh2022,feng2023metal,padilla2023metal,botzung2024,schlomer2024}.
In this paper, we have considered the related SU($N$) Holstein model, 
primarily for $N=3$.  
At half-filling we have shown that an insulating charge density wave phase
exists in which fully occupied sites of $N$ particles
alternate with empty sites, and we have determined the transition temperature as a
function of electron-phonon coupling.  
In that way, 
the ordering in the SU($N$) Holstein model appears to be more simple
than its Hubbard counterpart where distinct magnetic patterns can be
discerned at different commensurate fillings and even 
evolving with the interaction strength $U$.

The quintessential realization of SU($N$) symmetry in strongly
interacting quantum matter is accomplished with optically trapped alkaline earth atoms
and dipolar molecules\cite{karman2018,valtolina2020,schindewolf2022,bigagli2024,mukherjee2025,mukherjee2025a,karman2025}.
Indeed, a recent study\cite{stepp2025trion} has employed quantum simulations to examine trion
formation similar to that found here in the 
{\it attractive} Hubbard Hamiltonian, to which the Holstein model
explored here maps in the anti-adiabatic limit of large $\omega_0$.

Achieving higher $N$ in solid state systems is more problematic.
In multiorbital materials, the presence of a Hund's rule $J$ favors high spin states
and competes with the Hubbard $U$, breaking
SU($N$) symmetry.  
Exact SU(4) symmetry within the context of the Hubbard model has been proposed
for multi-orbital systems with fine-tuning of 
parameters\cite{gresista2023spin}, along with the
suggestions of materials, e.g.~$\alpha$-ZrCl$_3$
\cite{yamada2018emergent}, and
LiNiO$_2$\cite{li98},
in which SU(N) symmetry can emerge.
Another recent realization of an interacting electron system with potential SU(4) symmetry is in a
twisted multilayer configuration at magic angles with extremely narrow bands
\cite{bistritzer2011moire,goerbig11,koshino2018maximally,zhang2021SU4}.
Details of the low temperature phase behavior of these systems
ultimately will be controlled by the breaking of precise SU(N) symmetry,
but establishing the properties of SU(N) tight-binding Hamiltonians will
nevertheless constitute an important starting point for their modeling.

A natural extension of the results reported here is to the doped system.
For $N=2$, an $s$-wave superconducting phase emerges\cite{bradley2021superconductivity}
away from half-filling. 
Pairing for general $N$ has also been analyzed.
In \cite{honerkamp2004ultracold,honerkamp2004BCS}, the $N=3$ attractive Hubbard model was considered
within a mean field treatment, with the central focus being
on situations where two of the species have a non-zero gap $\Delta_{12} \neq 0$, while
the third remains gapless, $\Delta_{13} = \Delta_{23}= 0$. 
A similar possibility was examined with a Gutzwiller projected BCS 
variational trial  wavefunction\cite{rapp2007,rapp2008}, where it was noted
that if the density of different species is the same, domains would form, between
which the paired species and species densities would vary.
A uniform superfluid could emerge with different
global chemical potential of the paired and unpaired species.

In the present Holstein model, one might look for similar condensation.
However, such off-diagonal long range order requires temperatures close to an 
order of magnitude lower
than the diagonal ones arising from CDW formation at half-filling, and a factor of 2-3 lower
than that needed for superconductivity in the attractive Hubbard model\cite{bradley2021superconductivity}.
This is likely a consequence of large polaronic mass-  for a pair
to move, it must first break and then reform on a neighboring site.  
In short, even for $N=2$ superconductivity in the 
Holstein model is challenging to observe except in the
anti-adiabatic limit $\omega_{0} \gtrsim 2\,t$ of the attractive Hubbard model.
Since the energy scale to break a pair is proportional to $N$,
we expect looking for this phenomenon for $N>2$ will be even more challenging.

Finally, 
can SU($N$) symmetry manifest in electron-phonon materials?
As with the Hubbard case, fine-tuning would be required.  Specifically,
if higher $N$ were achieved through multiple electronic bands,
it would be necessary that the electron-phonon couplings
$\alpha_{j}$ would have to be equal.


\bibliography{SUNHolstein}

\newpage

\end{document}